\documentclass[prb,showpacs,preprintnumbers,amsfonts,amssymb,amsmath,floats,twocolumn]{revtex4}
\usepackage{graphicx}
\usepackage{amssymb,amsmath, multirow, longtable, bm, color}
\newcommand{\be}{\begin{equation}}
\newcommand{\ee}{\end{equation}}
\newcommand{\ba}{\begin{eqnarray}}
\newcommand{\ea}{\end{eqnarray}}
\newcommand{\bi}{\begin{itemize}}
\newcommand{\ei}{\end{itemize}}
\newcommand{\ff}[1]{{\bm #1}}
\newcommand{\tr}{\mbox{tr}}
\newcommand{\Tr}{\mbox{Tr}}
\newcommand{\refeq}[1]{Eq.\ (\ref{eq:#1})}

\begin{document}
\title{Accessing thermodynamics from dynamical cluster-embedding approaches} 

\author{Gang Li}

\affiliation{Institute for Theoretical Physics and Astrophysics, 
University of W\"urzburg, Am Hubland, 97074 W\"uzburg, Germany}

\author{Werner Hanke}

\affiliation{Institute for Theoretical Physics and Astrophysics, 
University of W\"urzburg, Am Hubland, 97074 W\"uzburg, Germany}

\author{Alexei N. Rubtsov}

\affiliation{Department of Physics, Moscow State University, 119992 Moscow, Russia}

\author{Sebastian B\"ase}

\affiliation{
I. Institute for Theoretical Physics,
University of Hamburg, Jungiusstr. 9, 20355 Hamburg, Germany
}

\author{Michael Potthoff}

\affiliation{
I. Institute for Theoretical Physics,
University of Hamburg, Jungiusstr. 9, 20355 Hamburg, Germany
}


\begin{abstract}
Dynamical quantum-cluster approaches, such as different cluster extensions of the dynamical mean-field theory (cluster DMFT) or the variational cluster approximation (VCA), combined with efficient cluster solvers, such as the quantum Monte-Carlo (QMC) method, provide controlled approximations of the single-particle Green's function for lattice models of strongly correlated electrons. To access the thermodynamics, however, a thermodynamical potential is needed. We present an efficient numerical algorithm to compute the grand potential within cluster-embedding approaches that are based on novel continuous-time QMC schemes: It is shown that the numerically exact cluster grand potential can be obtained from a quantum Wang-Landau technique to reweight the coefficients in the expansion of the partition function. The lattice contributions to the grand potential are computed by a proper infinite summation over Matsubara frequencies. A proof of principle is given by applying the VCA to antiferromagnetic (short-range) order in the two-dimensional Hubbard model at finite temperatures.
\end{abstract} 
 
\maketitle

\section{Introduction}
A powerful strategy to tackle strongly correlated fermion systems is to start from a local perspective, i.e.\ to treat the strong local Coulomb interaction exactly for an isolated cluster (or atom) first, and to include the coupling between the clusters in a subsequent step.
Dynamical cluster-embedding approaches \cite{Hettler-1998,Lichtenstein-2000,Senechal-2000,Kotliar-2001,Potthoff-2003-1,Okamoto-2003} (for reviews see Refs.\ \onlinecite{Georges-1996,Maier-2005,Kotliar-2006,Tremblay-2006}) provide this in a self-consistent way. 
The central object is the cluster single-particle Green's function $\ff G'(\omega)$ or the cluster self-energy $\ff \Sigma(\omega)$. 
These are computed exactly for the isolated cluster in an external dynamical (i.e.\ frequency dependent) Weiss field.
The Weiss field mimics the effect of the cluster surrounding and is calculated self-consistently using the exact cluster quantities and Dyson's equation for the lattice problem in the thermodynamic limit. 
Thereby, one has access to the single-particle excitation spectrum, i.e.\ to the (inverse) photoemission cross section, as well as to different physical quantities that can be derived from this. 

The thermodynamics of the system, however, is governed by a thermodynamical potential, such as the grand potential $\Omega$, which is related to the cluster Green's function and self-energy in a dynamical cluster embedding approach via:
\begin{equation}\label{eq:omega}
  \Omega = \Omega' + \mbox{Tr} \ln \ff G - \mbox{Tr} \ln \ff G' \: .
\end{equation}
Here $\Omega'$ is the grand potential of the cluster reference system, and $\ff G$ the approximate Green's function of the original lattice model in the thermodynamic limit which is obtained from the lattice Dyson equation using the cluster self-energy.
The trace refers to both, spatial and temporal lattice degrees of freedom, i.e.\ involves sums over lattice sites and Matsubara frequencies.

Since dynamical cluster-embedding is a concept that directly works in the thermodynamic limit, one of the main intentions is to construct phase diagrams. 
There are several well-known situations where the sole knowledge of the Green's function is insufficient, and a thermodynamcial potential is required. 

A prime example is the Mott transition in the paramagnetic phase of the half-filled Hubbard model. \cite{Mot90}
Using a plaquette of four sites, cellular dynamical mean-field theory (C-DMFT) predicts a finite $U$-range where the metallic and the Mott insulating solutions are coexisting. \cite{PHK08}
While the boundaries of the coexistence region could be mapped out precisely by using continuous-time quantum Monte-Carlo (CT-QMC), \cite{Rubtsov-2005, Werner-2006-1} the actual trend of the line of first-order transitions has not yet been determined. 
Recent calculations within the variational cluster approximation (VCA) by using the Lanczos method indicate that the first-order line does not end in a second-order critical point at zero temperature. \cite{BKS+09}

Coexistence and competition of phases with different long-range or short-range order is also characteristic for transition-metal oxides and cuprate materials in particular.
Salient features of this physics are captured by C-DMFT and VCA calculations for the doped single-band Hubbard model. \cite{Lichtenstein-2000,Civ09,SLMT05} 
Phase separation in the doped Mott insulator at $T=0$, for example, is another prime example where the knowledge of a thermodynamical potential is necessarily required.
\cite{AA05,Aichhorn-2006-2}

Note that for cluster-embedding approaches based on the self-energy-functional theory \cite{Potthoff-2003,Potthoff-2003a} an efficient evaluation of the grand potential via Eq.\ (\ref{eq:omega}) is decisive not only in the final step but {\em during} the actual calculation for exploiting a variational principle of the form $\delta \Omega[\ff \Sigma] = 0$ to find the physical self-energy. 
Again, the computation of $\Omega[\ff \Sigma]$ essentially follows Eq.\ (\ref{eq:omega}).
Furthermore, away from the stationary point, metastable phases as well as precursors of stable phases can be made visible by looking at the functional $\Omega[\ff \Sigma]$ (see Refs.\ \onlinecite{Potthoff-2003a,Dahnken-2004,Aichhorn-2006-2} for examples).

The cluster size that is accessible using full diagonalization (ED) or the Lanczos method \cite{LG93} as ``cluster solvers'' is strongly limited. 
Hubbard clusters consisting of more than, say, 12 sites cannot be treated conveniently in this way for zero and for finite $T$. \cite{LIM08} 
Therefore, if moderately large clusters are needed, finite-temperature quantum Monte-Carlo approach (QMC) represents the method of choice. \cite{HF86,Rubtsov-2005,Werner-2006-1}
An important advantage of the stochastic method is that uncorrelated (``bath'') sites, which make up the dynamic Weiss field in the cluster-embedding context, can be attached to the cluster of correlated sites essentially without any extra numerical cost. At the same time the bath helps to attenuate the QMC sign problem.

As concerns the evaluation of Eq.\ (\ref{eq:omega}) and thus the accessibility of the system's thermodynamics, however, ED turns out to be much more convenient: 
$\Omega'$ can be obtained from the cluster many-body eigenenergies.
The trace in the third term on the r.h.s.\ can easily be put in a form which involves the poles and the zeros of the cluster Green's function $\ff G'$ only, as has been shown in Ref.\ \onlinecite{Potthoff-2003a}.
The lattice contribution via the trace in the second term on the r.h.s. can also be evaluated with this information at hand by means of the so-called $Q$-matrix technique. \cite{Aichhorn-2006-2}

On the other hand, using QMC, there are two main obstacles that prevent a straightforward evaluation of Eq.\ (\ref{eq:omega}):
(i) As the Monte-Carlo technique is designed to provide expectation values, the cluster grand potential $\Omega'$ cannot be computed directly at finite $T$ because of the entropy term. An alternative is to find $\Omega'$ from $d\Omega' = - S' dT - N' d\mu + D' dU$ by integrating the cluster double occupancies $D'$ over $U$ for fixed temperature $T$ and chemical potential $\mu$. This, however, requires simulations for a finite $U$ range. 
(ii) The evaluation of the traces must be performed differently since QMC does not provide the poles and weights of the one-particle excitations. 
Furthermore, the Green's functions $\ff G'$ and $\ff G$ are available on the imaginary Matsubara frequencies only.
Integration along the real axis (as used e.g.\ in Ref.\ \onlinecite{Dahnken-2004} for VCA studies) is thus not possible. Instead, a direct numercal summation over the Matsubara frequencies must be employed, similar to an integration along the imaginary frequency axis as described in Ref.\ \onlinecite{Senechal-2008} for the $T=0$ limit.

The purpose of the present paper is to demonstrate that these difficulties can be overcome. 
We suggest to employ the continuous-time quantum Monte-Carlo method (CT-QMC) \cite{Rubtsov-2005, Werner-2006-1} combined with a quantum version of the Wang-Landau algorithm. \cite{Wang-2001,Wang-2001(2)} This allows for a direct numerical estimate of the cluster partition function for finite temperatures. 
The sign problem remains unaffected.
We also discuss the evaluation of the traces in Eq.\ (\ref{eq:omega}) by summing over Matsubara frequencies. 
The paper is organized as follows: In Sec.\ \ref{WL-CTQMC} we show how to combine the quantum Wang-Landau algorithm with CT-QMC to determine the grand potential of an isolated cluster. For accuracy checks, the results are compared with those from full diagonalizations of small clusters. In Sec.\ \ref{Ther_Limit} the Matsubara-frequency summation is discussed emphasizing the analytical treatment of the high-frequency limit. The application of the technique to the 2D square Hubbard system by embedding a $4\times 4$ cluster in an infinite square lattice is demonstrated in Sec.\ \ref{application}, where various thermodynamical properties are discussed. 

\section{Quantum Wang-Landau approach}\label{WL-CTQMC}

Usual Monte-Carlo algorithms sample configurations in the configuration space by an ergodic random walk. In practice, however, the Monte-Carlo walk could be trapped in a certain part of the configuration space, especially when the system is close to a discontinuous phase transition.
The Wang-Landau algorithm \cite{Wang-2001,Wang-2001(2)} has been introduced to overcome such problems in the classical systems. For quantum systems, it has been proposed for an efficient sampling in the context of a stochastic series expansion in the inverse temperature. \cite{Troyer-2003} Basically the same idea can be applied to algorithms with different expansion parameters. Here we will discuss the application of Wang-Landau algorithm in the context of the continuous-time quantum Monte-Carlo method \cite{Rubtsov-2005, Werner-2006-1, Assaad-2007, Werner-2006-2} where the partition function of a fermionic quantum system is expanded in powers of the interaction (or hybridization) strength.

To discuss this quantum Wang-Landau approach, we refer the so-called "weak-coupling" CT-QMC \cite{Rubtsov-2005} as an example.
We also consider the single-band Hubbard model which reads
\begin{equation}\label{Eq:Hubbard}
H = -t\sum_{\langle i,j\rangle,\sigma}(c_{i\sigma}^{\dagger}c_{j\sigma} + c_{j\sigma}^{\dagger}c_{i\sigma}) - \mu\sum_{i}n_{i} + U\sum_{i}n_{i\uparrow}n_{i\downarrow} \: ,
\end{equation}
in the usual notation with $c_{i\sigma}^\dagger$ being the creation operator of an electron at site $i$ and spin projection $\sigma=\uparrow,\downarrow$, with $n_{i\sigma}=c_{i\sigma}^\dagger c_{i\sigma}$, the nearest-neighbor hopping parameter $t$, the chemical potential $\mu$ and the interaction strength $U$. 
As a non-perturbative and numerically exact method, the CT-QMC starts from the infinite sum over diagram orders $k$ in the expansion of the partition function: 
\begin{equation}\label{Eq:CT-QMCExpansion}
  \frac{{\cal Z}}{{\cal Z}_{0}} = \sum_{k=0}^{\infty} U^k w(k) = 1 + \sum_{k=1}^{\infty} U^k w(k) \: ,
\end{equation}
where ${\cal Z}_{0}$ is the partition function of the non-interacting system. 
The coefficient of the $k$-th order
\begin{equation}\label{Eq:weights}
  w(k) = \sum_{C_k} w(k,C_k) 
\end{equation}
is given as a sum over vertex configuration $C_k$ ($k$ vertices at order $k$) which specifies the positions of the vertices in space and imaginary time.
It includes a $k$-dimensional $\tau$-integration over $[0,\beta]$.
The weights
\begin{equation}\label{Eq:weights1}
  w(k,C_k) = (-\Delta \tau/2)^k \det M_{\uparrow}^{(k)}(C_k) \det M_{\downarrow}^{(k)}(C_k)
\end{equation}
are composed of determinants of matrices $M_\sigma^{(k)}(C_k)$ which are constructed from non-interacting Green's functions that link the $k$ vertices of a vertex configuration $C_k$.
A finite $\Delta \tau$ is used for a discretization of the $\tau$-integrations but can be made arbitrarily small.
This sum over vertex configurations is high-dimensional for all practical purposes. 
For details of the CT-QMC method see Refs.\ \onlinecite{Rubtsov-2005,Assaad-2007}, for example. 

A fermion sign problem can be avoided for the model at half-filling using a parameter $\alpha$, \cite{Rubtsov-2005} in the interaction term, 
\begin{equation}
H_{U}=\frac{U}{2}\sum_{i}[(n_{i\uparrow}-\alpha)(n_{i\downarrow}-1+\alpha) + (n_{i\uparrow}-1+\alpha)(n_{i\downarrow}-\alpha)] \; ,
\end{equation}
and in the one-particle part 
\begin{equation}
H_{0} = -t\sum_{\langle i,j \rangle,\sigma}(c_{i\sigma}^{\dagger}c_{j\sigma} + c_{j\sigma}^{\dagger}c_{i\sigma} ) - (\mu-U/2)\sum_{i}n_{i}  - N(\alpha-\alpha^{2})U\: .
\end{equation}
Here $N$ is the total number of sites in the system. 
The non-interacting partition function can be determined easily by diagonalizing $H_{0}$. This yields:
\begin{equation}
{\cal Z}_{0}  = e^{\beta N(\alpha-\alpha^{2})U}{\mbox{Tr}} \: e^{-\beta\sum_{q\sigma}(\varepsilon_{q\sigma}-\mu + U/2)n_{q\sigma}} \: ,
\end{equation}
where $q$ is the momentum vector. It can be further written as
\begin{equation}
{\cal Z}_{0} = e^{\beta N(\alpha-\alpha^{2})U}\prod_{q}[1+e^{-\beta(\varepsilon_{q}-\mu+U/2)} ]^{2} \: .
\end{equation}

\begin{figure}[htbp]
\centering
\includegraphics[width=\linewidth]{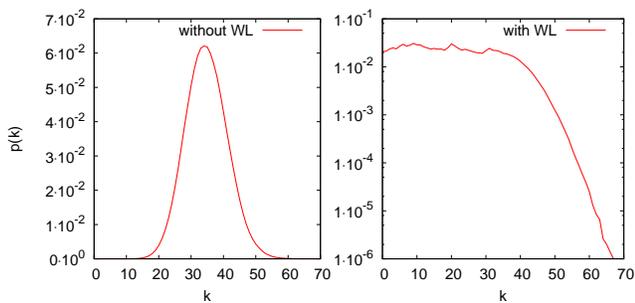}
\caption{Left: Histogram $p(k)$ of the order $k$ in the weak-coupling expansion.
Calculation for a $2\times 2$ Hubbard cluster at half-filling, $\beta t = 5$ and $U/t = 4$. Right: Histogram with Wang-Landau re-weighting, $\tilde{p}(k)$, choosing $k_c = 40$.
}\label{Fig:Histogram}
\end{figure}

The high-dimensional sum over orders $k$ and vertex configurations $C_k$ is sampled by a Monte-Carlo technique.
Thereby, one can deduce the weight function $w(k)$ up to a constant factor.
Note that $w(k)$ is not normalized to unity (Eq.\ (\ref{Eq:CT-QMCExpansion}) yields
$\sum_{k=0}^\infty w(k)= {\cal Z}_{U=1}/{\cal Z}_{0}$). 
A corresponding histogram,
\begin{equation}
p(k) = U^k w(k) \: ,
\end{equation}
generated by CT-QMC for a small ($2\times 2$) Hubbard cluster, is shown in the left part of Fig.\ \ref{Fig:Histogram}. 
Since $w(0) = p(0) = 1$, as is obvious from Eq.\ (\ref{Eq:CT-QMCExpansion}), one can in principle determine the unknown factor from the $k=0$ term in the histogram and find the partition function as ${\cal Z} = {\cal Z}_{0}[1 + \sum_{k=1}^{\infty} p(k)/p(0)]$. However, this is by no means practicable, since usually $p(0)$ is negligibly small compared to $p(k)$ at the average order, for example (see Fig.\ \ref{Fig:Histogram}, left).
Therefore a histogram re-weighting technique is necessary. 

As discussed in Ref.\ \onlinecite{Troyer-2003}, the basic idea of Wang-Landau sampling is to create a histogram $\tilde{p}(k)$ which is flat for all orders $k$ up to a certain cutoff order $k_{c}$.
Thereby, the algorithm generates approximately the same number of configurations $C_k$ both, at low orders and at higher orders (up to $k_c$).
To achieve this, a Wang-Landau factor $g(k)$ is introduced to re-define the weights $w(k,C_k)$:
\begin{equation}\label{eq:WLg}
w(k,C_k) \to \widetilde{w}(k,C_k) = w(k,C_k) / g(k) \: .
\end{equation}
This also implies the replacement
\begin{equation}
w(k) \to \widetilde{w}(k) = w(k) / g(k) \: .
\end{equation}
The (e.g.\ Metropolis) random walk is performed in the usual way, but with respect to the new weights $\tilde{w}(k,C_k)$ and thereby with new transition probabilities. The Wang-Laudau factor $g(k)$ is chosen to make the new histogram flat, i.e.\ $\tilde{p}(k)=U^k\tilde{w}(k)=const$ for $k<k_{c}$. 
Hence, the partition function is then given by 
\begin{equation}\label{eq:xxx}
\frac{\cal Z}{{\cal Z}_{0}} = \sum_{k=0}^{\infty} U^k w(k) 
= \sum_{k=0}^{\infty} U^k \tilde{w}(k) g(k) \: .
\end{equation}
If the histogram was completely flat to all orders, 
\begin{equation}
\frac{\cal Z}{{\cal Z}_{0}} 
= \tilde{p}(0) \sum_{k=0}^{\infty} g(k) \: ,
\end{equation}
and with $p(0)=1$ we would get
\begin{equation}
{\cal Z}/{\cal Z}_{0} = \sum_{k=0}^{\infty} g(k)/g(0) \: .
\end{equation}
Note, however, that in practice the Wang-Landau re-weighting is performed up to certain order $k_{c}$ only. 
Furthermore, in any practical simulation $\tilde{p}(k)=const$ is approximate (but can be ensured to arbitrary precision in principle).
Therefore, Eq.\ (\ref{eq:xxx}) with the actual new weights and the actual Wang-Landau factor has to be used for concrete calculations.
The cutoff order $k_{c}$ is basically a preselected number representing up to which order the histogram will be re-weighted. 
Typically, $k_c > \langle k \rangle$, where $\langle k \rangle$ is the average order, has turned out to be a good choice.

As discussed in Refs.\ \onlinecite{Wang-2001,Troyer-2003}, the Wang-Landau factor $g(k)$ is continuously modified during a simulation by multiplying $g(k)$ with a constant $f$ if a configuration at order $k$ is visited,
\begin{equation}
g(k) \to g(k) \cdot f \; ,
\end{equation}
until the measured histogram $\tilde{p}(k)$ is flat. 
In practice, we set $g(k)=e^{G(k)}$, and increase $G(k)$ at each visit by $F$, i.e. $G(k) \to G(k) + F$. 
In our implementation of the algorithm, the flatness of $\tilde{p}(k)$ is defined by requiring the difference of the smallest and largest $\tilde{p}(k)$ to be within a given small tolerance $\eta>0$, i.e.
\begin{equation}\label{Eq:flat}
\tilde{p}_{min} \ge (1-\eta) \tilde{p}_{max} \: .
\end{equation}
In the actual simulation, multiple steps of Wang-Landau re-weighting is recommended. Each time when the above condition is fulfilled, a further decrease of the multiplicative constant $f$ improves the flatness of the histogram.  

The Wang-Landau factor $g(k)$ is positive by construction. This implies that the new weights $\widetilde{w}(k,C_k)$ are positive if and only if the original weights $w(k,C_k)$ are positive. Therefore, the re-weighting technique does not introduce a new source for a sign problem.

Fig.\ \ref{Fig:Histogram} (right) shows the re-weighted histogram. We can see that after the Wang-Landau re-weighting up to $k_c=40$, the histogram $\tilde{p}(k)$ is sufficiently ``flat''. 
Each time when the criterion Eq.\ (\ref{Eq:flat}) is fulfilled, $F$ has been reduced by a factor two, and the tolerance decreased by $0.05$ to improve the flatness of the histogram.
For the example shown and also for all calculations below $F=0.01$ and a tolerance $\eta=0.2$ as final values have turned out to be sufficient.
Note that if ${\cal Z}/{\cal Z}_{0}$ is computed from Eq.\ (\ref{eq:xxx}) it is unnecessary to have a {\em strictly} flat histogram.

\begin{figure}[htbp]
\centering
\includegraphics[width=\linewidth]{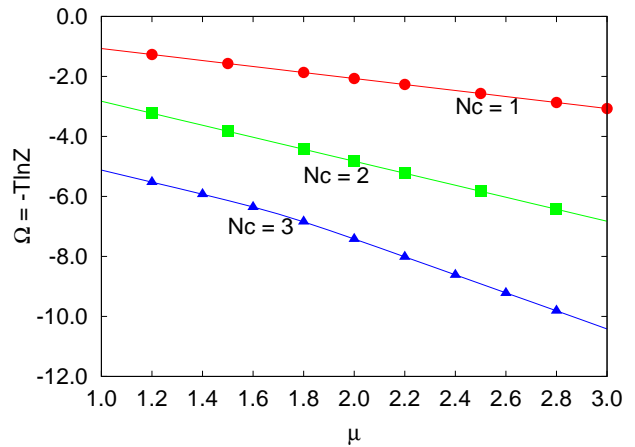}
\caption{The grand potential per site as a function of the chemical potential $\mu$ for small Hubbard clusters at $\beta t=10$ and $U/t=4$. The CT-QMC results are shown as dots. The statistical error is smaller than the symbol size. Full ED results are shown as lines for comparison.}\label{Fig:PartitionFunction}
\end{figure}

In Fig.\ \ref{Fig:PartitionFunction}, the grand potential $\Omega=-T\ln{\cal Z}$ is shown for three different cluster sizes. 
$\Omega$ is calculated as a function of the chemical potential at $\beta t=10$ and $U/t=4$ and compared to numerically exact results from full diagonalization of the problem. 
The agreement with the exact grand potential is excellent. The standard deviation of the cluster grand potential is reasonably small in our calculation, which is crucial since the lattice grand potential consists of the cluster grand potential $\Omega^{\prime}$ and of $\mbox{Tr} \ln \ff G - \mbox{Tr} \ln \ff G^{\prime}$. Both contributions are of the same order of magnitude. 
Hence an accurate determination the cluster grand potential is important. Within the Wang-Landau CT-QMC implementation, we find that the statistical error of $\Omega^{\prime}$ is comparable to that of the $\mbox{Tr} \ln \ff G - \mbox{Tr} \ln \ff G^{\prime}$, and typically three orders of magnitude lower than its mean value for the calculations presented here.

Besides the possibility to compute the grand potential, there is another important advantage of the quantum Wang-Landau method:
Suppose that we have the Wang-Landau factor $g(k)$ from a calculation at interaction strength $U_0$. 
Then the full $U$ dependence of the average of an observable can be determined at once for $0\le U \le U_0$ since the $U$ dependence of the weights is trivially $U^k w(k,C_k)$. 
Since $w(k,C_k)$ is a functional of the free Green's function and $\beta$ only, it is $U$ independent. 

The average of an observable ${\cal O}$ is
\begin{equation}
\langle {\cal O} \rangle = 
\frac{\sum_{k} \sum_{C_k} U^k w(k,C_k) {\cal O}(k,C_k)}{\sum_{k} \sum_{C_k} U^k w(k,C_k)} \: .
\end{equation}
Using Eq.\ (\ref{eq:WLg}), we have
\begin{equation}
\langle {\cal O} \rangle = 
\frac{\sum_{k} U^k g(k) \sum_{C_k} \tilde{w}(k,C_k) {\cal O}(k,C_k)}{\sum_{k} U^k g(k) \sum_{C_k} \tilde{w}(k,C_k)} \: .
\end{equation}
Define an average at a given order $k$ by
\begin{equation}
{\cal O} (k) =
\frac{\sum_{C_k} \tilde{w}(k,C_k) {\cal O}(k,C_k)}{\sum_{C_k} \tilde{w}(k,C_k)} \: .
\end{equation}
This average is carried out by importance sampling of configurations $C_k$ in the Monte-Carlo technique.
The subsequent average over different perturbation orders $k$ then reads as
\begin{equation}\label{Eq:UInterpolation}
\langle {\cal O} \rangle 
= \frac{\sum_{k} U^k g(k) {\cal O} (k) }{\sum_{k} U^k g(k)} \: .
\end{equation}
The $U$ dependence of $\langle {\cal O} \rangle$ is now explicit. 

The $U$-dependent average perturbation order $\langle k \rangle$, the average particle numbers, the interacting Green's function, etc.\ are easily obtained from this equation. Note that since $U/U_0\le 1$, only smaller number of diagrams is required in the construction of the full partition function at $U$ compared to $U_0$, while $g(k)$ is determined up to the maximum contributing order at $U_0$. 

\begin{figure}[htbp]
\centering
\includegraphics[width=\linewidth]{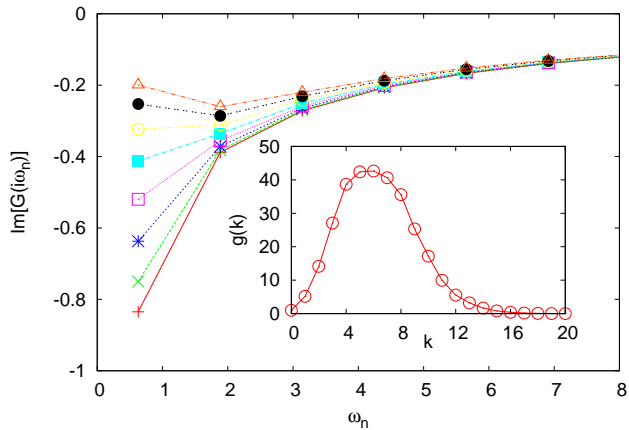}
\caption{The on-site Green's function for a $2\times 2$ cluster at $\beta t=5$. From bottom to top, different lines corresponds to different interaction strengths varying from $U/t=0.5$ to $U/t=4.0$ in steps of $0.5$. The Wang-Landau factor $g(k)$ was determined at $U/t=4$, and the Green's function for other interaction strengths are calculated from Eq.\ (\ref{Eq:UInterpolation}).}
\label{Fig:GreensFunction}
\end{figure}

Fig.\ \ref{Fig:GreensFunction} shows the Wang-Landau factor $g(k)$ at $\beta t=5$ and $U/t =4$ on a 2$\times$2 cluster as well as the on-site Green's functions for $U=0.5$ to $U=4$ with steps $\Delta U=0.5$ as functions of the Matsubara frequency. 
The Green's functions have been calculated using Eq.\ (\ref{Eq:UInterpolation}).
We have checked that these results agree with those from conventional CT-QMC simulations at the respective fixed $U$. The low-frequency behavior is reminiscent of the metal-insulator transition in the infinite square lattice.

\section{Lattice contribution to the grand potential}\label{Ther_Limit}

Within a cluster-embedding approach, the approximate grand potential of the lattice fermion model in the thermodynamical limit with parameters $\ff t$ and $U$ is given via \refeq{omega} by the grand potential $\Omega'$ of a small cluster with parameters $\ff t'$ and $U$ and a lattice contribution. 
The latter is obtained from the approximate single-particle lattice Green's function,  ${\ff G}_{{\ff t}, U}^{-1} = i\omega_{n}+\mu-{\ff t}-{\ff \Sigma}_{{\ff t^{\prime}}, U}$, and the exact cluster Green's function,
${\ff G}_{{\ff t^{\prime}}, U}^{-1} = i\omega_{n}+\mu-{\ff t^{\prime}}-{\ff \Sigma}_{{\ff t^{\prime}}, U}$, as
\begin{eqnarray}\label{Eq:diff}
&&\Tr\ln({\ff G}_{{\ff t}, U}) - \Tr\ln({\ff G}_{{\ff t^{\prime}}, U})\nonumber\\
&=&2T\sum_{n,\sigma}e^{i\omega_{n}0^{+}}\frac{N_{c}}{N}\sum_{\ff{\tilde k}}\tr\ln\frac{{\ff G}_{\ff t^{\prime}, U}(\ff{\tilde k},i\omega_{n})}
{{\ff G}^{\prime}_{\ff t^{\prime}, U}(i\omega_{n})}\nonumber\\
&=&-2T\frac{N_{c}}{N}\sum_{n,\sigma, {\ff{\tilde k}}}e^{i\omega_{n}0^{+}}\ln\det[1-{\ff V_{\tilde{k}}}{\ff G}_{{\ff t^{\prime}}, U}(i\omega_{n})] \: .
\end{eqnarray}
Here $\Tr=2T\sum_{\ff{\tilde k}}\sum_{n}e^{i\omega_{n}0^{+}}{\rm tr}$ combines the usual trace referring to the cluster sites, the sum over Matsubara frequencies, and the sum over wavevectors in the reduced Brillouin zone of the superlattice of disconnected clusters. 
The factor 2 accounts for the two spin projections.
Furthermore, ${\ff V}_{\tilde{k}} = {\ff t}_{\ff{\tilde k}} - {\ff t^{\prime}}$ is the hybridization matrix stressing the difference between the lattice and the cluster system with respect to the one-particle parameters. $N_{c}$ and $N$ are the size of the cluster and of the  lattice, respectively.

There are in principle two different techniques that can be employed to evaluate the frequency sum in Eq. (\ref{Eq:diff}): (i) An analytical evaluation is possible, once the poles and the weights of the cluster Green's function are available. This is the idea of the $Q$-matrix method \cite{Potthoff-2003a,Aichhorn-2007-2} which works well at zero temperature where only a few number of poles contribute, and for small clusters where the full eigensystem is available from exact diagonalization. It is obvious, however, that the CT-QMC technique is not capable to locate the poles. (ii) Alternatively, one can evaluate the frequency summation numerically using the CT-QMC result for ${\ff G}_{{\ff t^{\prime}}, U}(i\omega_{n})$ on the Matsubara frequencies. This summation can be performed in close analogy to the integration along the imaginary frequency axis described by S\'{e}n\'{e}chal. \cite{Senechal-2008} 

Below we briefly describe the method that has been used here:
Note that a direct numerical frequency summation, in Eq.\ (\ref{Eq:diff}), is not possible since the factor $e^{i\omega_{n}0^{+}}$ is crucial for convergence.
One therefore has to treat the high-frequency part separately and analytically.
Introducing a (sufficiently large) cutoff frequency $\omega_{\Lambda}$, the infinite Matsubara sum can be split up in the following way:
\begin{eqnarray}\label{Eq:summ}
&&\Tr\ln({\ff G}_{{\ff t}, U}) - \Tr\ln({\ff G}_{{\ff t^{\prime}}, U}) \nonumber\\
&=&-2T\sum_{{\ff{\tilde k}}}\frac{N_{c}}{N}\sum_{n=-\Lambda}^{\Lambda}e^{i\omega_{n}0^{+}}\ln\frac{\det\left[1-{\ff V_{\tilde{k}}}{\ff G}_{{\ff t^{\prime}}, U}(i\omega_{n})\right]}
{\det\left[1-{\ff V_{\tilde{k}}}/i\omega_{n}\right]}\nonumber\\
 &&-2T\sum_{\ff{\tilde k}}\frac{N_{c}}{N}\sum_{n=-\infty}^{\infty}e^{i\omega_{n}0^{+}}\ln\det\left[1-\frac{\ff V_{\tilde{k}}}{i\omega_{n}}\right]
\end{eqnarray}
The first term involves a finite Matsubara sum only and can thus be computed by direct numerical summation with $e^{i\omega_{n}0^{+}}$ set to unity. 
For the second term, the frequency summation can be done analytically. We find:
\begin{eqnarray}\label{Eq:relation0}
&&-2T\sum_{\ff{\tilde k}}\frac{N_{c}}{N}\sum_{n=-\infty}^{\infty}e^{i\omega_{n}0^{+}}\ln\det\left[1-\frac{\ff V_{\tilde{k}}}{i\omega_{n}}\right] \nonumber\\
&=&2T\left[N_{c}\ln2 - \sum_{a, \ff{\tilde k}}\frac{N_{c}}{N}\ln(1+ e^{-{\ff{\tilde V}^{aa}_{\tilde{k}}/T}})\right]
\end{eqnarray}
where ${\ff{\tilde V}^{aa}_{\tilde{k}}}$ are the diagonal elements of ${\ff V_{\tilde{k}}}$ and $a$ refers to the sites in the cluster.
This yields
\begin{eqnarray}\label{Eq:final}
\Omega &=& \Omega^{\prime}
-2T\sum_{{\ff{\tilde k}}}\frac{N_{c}}{N}\sum_{n=-\Lambda}^{\Lambda}\ln\frac{\det\left[1-{\ff V_{\tilde{k}}}{\ff G}_{{\ff t^{\prime}}, U}(i\omega_{n})\right]}
{\det\left[1-{\ff V_{\tilde{k}}}/i\omega_{n}\right]}\nonumber\\ 
& & + 2TN_{c}\ln2 - 2T\sum_{a, \ff{\tilde k}}\frac{N_{c}}{N}\ln(1+ e^{-{\ff{\tilde V}^{aa}_{\tilde{k}}/T}})
\end{eqnarray}
for the lattice grand potential of a cluster-embedding approach at finite temperatures. 
Within CT-QMC, the cluster contribution $\Omega^{\prime}$ is determined as shown in Sec.\ \ref{WL-CTQMC}. Note that, for different cluster-embedding methods, different implementations of the non-interacting partition function ${\cal Z}_{0}$ must be considered (see Ref.\ \onlinecite{GangLi-2009-1}). 

The one-particle Green's function ${\bf G}_{\ff t^{\prime}, U}(i\omega_{n})$ is measured up to the cutoff frequency $\omega_{\Lambda}$.
Clearly, the accuracy of the frequency summation in the lattice contribution to the grand potential is sensitive to this cutoff. 
To check the accuracy, we consider as a simple test case the half-filled Hubbard model with a semi-elliptical free density of states $\rho_0(z)$ of bandwidth $W=4t$, and the Hubbard atom ($N_c=1$) without bath sites as a reference ``cluster''. The cluster Green's function at half-filling is readily obtained, the corresponding local self-energy is ${\Sigma}_{\ff t^{\prime}, U}(i\omega_{n}) = U^{2}/4i\omega_{n}$.
Within the cluster-embedding method, this approximates the self-energy of the lattice model and yields the lattice Green's function, the local elements of which are thus given in terms of the free density of states $\rho_0(z)$ as
\begin{equation}
{G}_{\ff t, U}(i\omega_n) = \int_{-W/2}^{W/2}\frac{\rho_0(z)dz}{i\omega_{n} - z - {\Sigma}_{\ff t^{\prime}, U}(i\omega_{n})} \: .
\end{equation}
The poles of lattice Green's function are determined by solving the equation $i\omega_{n} - z - {\bf \Sigma}_{\ff t^{\prime}, U}(i\omega_{n})=0$. 
From these and from the poles of ${\bf G}_{\ff t^{\prime}, U}(i\omega_{n})$ the lattice grand potential can be determined analytically. We find:
\begin{eqnarray}\label{Eq:analy}
\Omega &=& \Omega^{\prime} + T\ln\left[1+\cosh(\frac{U}{2T})\right]\nonumber\\  &-&T\int\rho_0(z)dz\ln\left[\cosh\frac{z}{2T}+\cosh\frac{\sqrt{z^{2}+U^{2}}}{2T}\right] \: .
\end{eqnarray}
This is easily evaluated numerically to any desired accuracy by means of the Simpson method.
Table \ref{Table:compare} compares the exact result for the lattice grand potential (per site) with the numerical result obtained by evaluation of Eq. (\ref{Eq:final}) for different cutoff frequencies. 
This demonstrates that a moderate cutoff frequency is sufficient to obtain a reliable result with a relative error, as compared to the exact result from Eq. (\ref{Eq:analy}), of the order of $10^{-3}$.
This improves with increasing cutoff frequency.
With decreasing temperature, the required frequency cutoff becomes larger.

\begin{table}[htdp]
\begin{center}
\begin{tabular}{| c || c | c | c |}
\hline
$\beta$ & Exact & \multicolumn{2}{| c |}{Numerical: Eq. (\ref{Eq:final})} \cr
\hline
\multirow{2}{*}{1.0} & \multirow{2}{*}{-0.03423} & $\Lambda = 10$ & $\Lambda = 30$ \cr
\cline{3-4}
& & -0.03425 (5.8E-4)& -0.03424 (3.0E-4) \cr
\hline
\multirow{2}{*}{10.0} & \multirow{2}{*}{-0.03099} & $\Lambda = 50$ & $\Lambda = 100$ \cr
\cline{3-4}
& & -0.03106 (2.3E-3)& -0.03102 (9.7E-4) \cr
\hline
\end{tabular}
\end{center}
\caption{
Grand potential per site for the Hubbard model at half-filling with semi-elliptical free density of states (bandwidth $W=4t$) as obtained for a reference ``cluster'' with $N_c=1$. Numerical results obtained from Eq.\ (\ref{Eq:final}) for different temperatures and cutoff frequencies are compared to the exact result (Eq.\ (\ref{Eq:analy})). The relative differences are given in brackets. \label{Table:compare}}
\end{table}%

\section{Short-range antiferromagnetic order at finite temperatures}\label{application}

For an application of CT-QMC combined with the quantum Wang-Landau method we consider the Hubbard model on the square lattice at half-filling and $\beta t \le 5$ and use a $4\times4$ Hubbard cluster for the embedding. This is well beyond the cluster size that can be accessed (conveniently) by exact diagonalization or Lanczos within a cluster-embedding approach and thus sufficient to provide a proof of principle. Clearly, our method to determine a thermodynamical potential can easily be extended to study larger clusters and lower $T$ with more computational effort. 

Generally, the additional computational effort necessary for the reweighting technique is reasonable: First, we note that the time needed to initially determine the Wang-Landau factor is usually less than the time spent for the measurement of observables in our calculations.
It is even negligible in those cases where there is a good initial estimate for the Wang-Landau factor, as e.g.\ from a preceeding calculation with slightly different model parameters. Second, since the reweighting technique amounts to importance sampling for any given order $k$ but full sampling over all $k$ up to the cutoff order, the configuration space is enlarged considerably. However, the orders to be sampled additionally are computationally inexpensive because they are significantly lower than the average order. The numerical effort is still dictated by the average order as in conventional CT-QMC.

\begin{figure}[htbp]
\centering
\includegraphics[width=\linewidth]{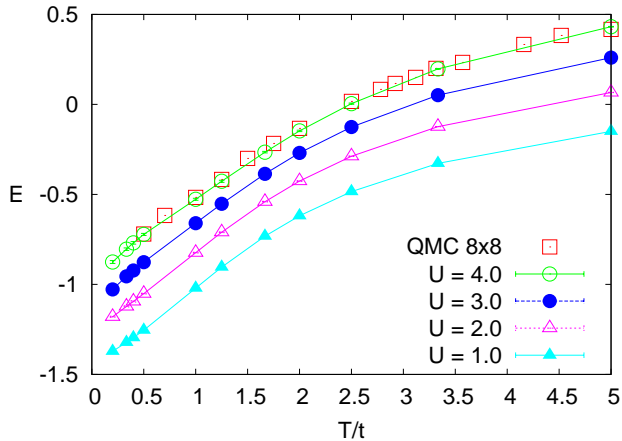}
\caption{The internal energy per site determined from a $4\times4$ embedded cluster at different temperatures. From bottom to top, the interaction strength increases from $U/t=1.0$ to $4.0$. Statistical error bars are smaller than the symbol size and hardly visible. Our results are in good agreement with the QMC solution on an $8\times8$ cluster obtained in Ref.\ \onlinecite{Duffy-1995} at $U/t = 4.0$.}\label{InternalEnergy}
\end{figure}

To keep things simple we furthermore use clusters without bath sites, i.e.\ we perform a VCA calculation where a single variational parameter is considered only. Note that bath sites could be added without any additional numerical cost within the weak-coupling CT-QMC. For the particle-hole symmetric model at half-filling, there is no sign problem.

The strength $h$ of a staggered magnetic Weiss field suggests itself as the most relevant variational parameter. Hence, we add the term
\begin{equation}\label{eq:weiss}
 H_{\rm Weiss} = h\sum_{i}e^{i{\bf Q}\cdot {\bf r}_{i}}(n_{i\uparrow}-n_{i\downarrow})
\end{equation}
to the Hamiltonian of the reference cluster. Here, ${\bf Q}=(\pi, \pi)$ is the antiferromagnetic wave vector. 
To get the optimal value of the Weiss field $h$ within the framework of the self-energy-functional approach, we compute the lattice grand potential $\Omega$ for different $h$ according to the method outlined in the previous sections and search for stationary points:
\begin{equation}
\frac{\partial \Omega}{\partial h} = 0 \: .
\end{equation}
This optimization of the thermodynamical potential can also be performed in the presence of a finite {\em physical} external staggered field of strength $h_{\rm AF}$ which must not be confused with the variational paramerter $h$. The AF order parameter, i.e.\ the staggered magnetization, is then obtained as
\begin{equation}\label{eq:m}
m = \lim_{h_{\rm AF}\rightarrow 0}\frac{\partial{\Omega}}{\partial h_{\rm AF}} \: .
\end{equation}
Spontaneous symmetry breaking is indicated by a finite $m$, or, looking at $\Omega(h)$ for $h_{\rm AF}=0$, by a finite optimal value for the variational parameter $h$. 

\begin{figure}[htbp]
\centering
\includegraphics[width=\linewidth]{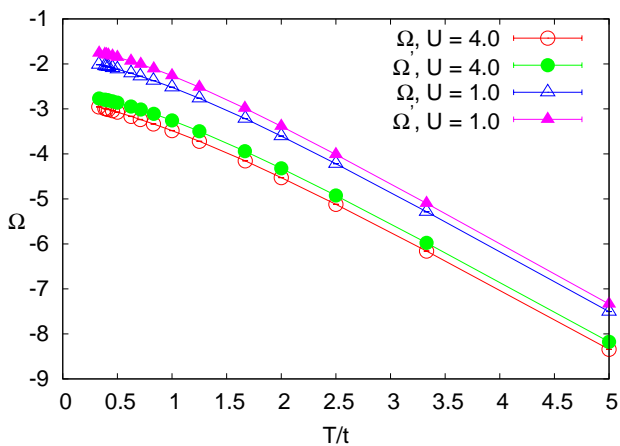}
\caption{The grand potential $\Omega$ per site from Eq. (\ref{Eq:final}) and the cluster grand potential $\Omega'$ per site as functions of temperature for two different interaction strengths.}\label{GrandPotential}
\end{figure}

For our calculations we usually re-weighted the histograms up to the order $k_{c} = N_{c}\beta U/2$. For  higher temperatures, it turned out to be useful to extend $k_c$ by 20 more orders for the re-weighting.  The Wang-Landau factor $g(k)$ was kept fixed when the histogram for $k<k_{c}$ became sufficiently flat, as controlled by Eq.\ (\ref{Eq:flat}). 

Each observable was measured by totally $2.56\times10^{8}$ Monte Carlo steps after a flat histogram has been obtained, see Eq. (\ref{Eq:flat}). This turned out to be sufficient for controlling the statistical error.
In most of the results shown below, error bars are smaller than the symbol size. The frequency summations discussed in Sec.\ \ref{Ther_Limit} have been carried out with a frequency cutoff of $\Lambda = 120$ for all temperatures considered here. 

\begin{figure}[htbp]
\centering
\includegraphics[width=\linewidth]{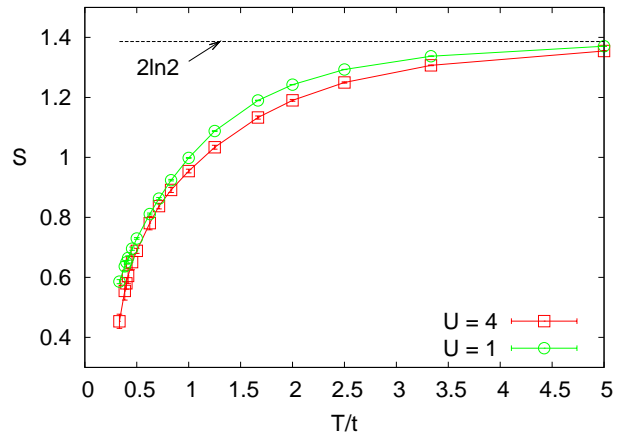}
\caption{Entropy (per site) as calculated from Eq.\ ({\ref{Eq:Entropy}}) for the same parameters as for Fig. \ref{InternalEnergy} and \ref{GrandPotential}.}\label{Entropy}
\end{figure}

To start with, we discuss results for vanishing Weiss field, i.e. $h=0$. Fig.\ \ref{InternalEnergy} shows the internal energy as a function of temperature for different interaction strengths. The internal energy is computed as $E = E_{kin} + U\langle D\rangle$ from the double occupancy $\langle D\rangle$ and the kinetic energy $E_{kin}$ which is obtained from the lattice Green's function (i.e.\ with self-energy replaced by the cluster self-energy). \cite{Georges-1996} The agreement of our results at $U/t = 4.0$ with those of a QMC calculation \cite{Duffy-1995} for an isolated Hubbard cluster of $8\times8$ sites is excellent in the entire temperature range. 

While the internal energy (as an expectation value of an observable) is directly available within usual QMC methods, a thermodynamical potential is not. The advantage of the Wang-Landau approach lies in the direct accessibility of the partition function and thus of the grand potential for any $U\le U_0$ -- if implemented within the (weak-coupling) CT-QMC framework. 

Fig.\ \ref{GrandPotential} shows the temperature dependence of the grand potential $\Omega$ as obtained from Eq.\ (\ref{Eq:final}) for two different interaction strengths. 
The cluster grand potential $\Omega^{\prime} = -T\ln{\cal Z}^{\prime}$ with ${\cal Z}^{\prime}$ determined by the  Wang-Landau approach is shown in addition.
Using the $4\times 4$ cluster, $\Omega'$ already represents the main contribution to the total grand potential. 
As a function of temperature, the lattice contribution $\Tr \ln \ff G - \Tr \ln \ff G'$ resulting from the cluster embedding is basically constant at high temperatures but is of crucial importance for low $T$.
This has to be expected as spatial correlations grow with decreasing $T$.

\begin{figure}[htbp]
\centering
\includegraphics[width=\linewidth]{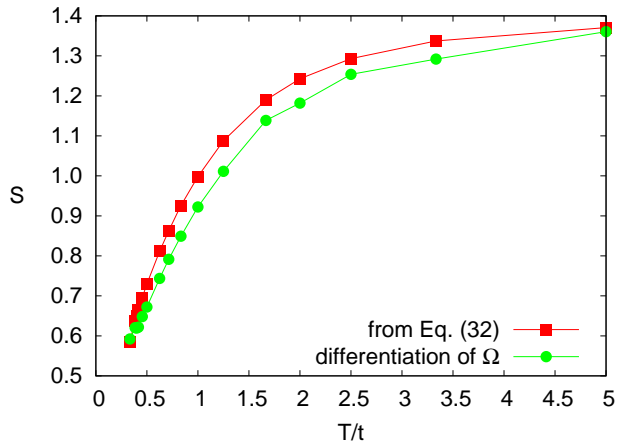}
\caption{The comparison of the entropy from Eq. (\ref{Eq:Entropy}) and the numerical differentiation of the the lattice grand potential, $S = - \partial \Omega / \partial T$, at $U/t=1$ for different temperatures.}\label{Fig:Compare_S}
\end{figure}

\begin{figure*}[htbp]
\includegraphics[width=\textwidth]{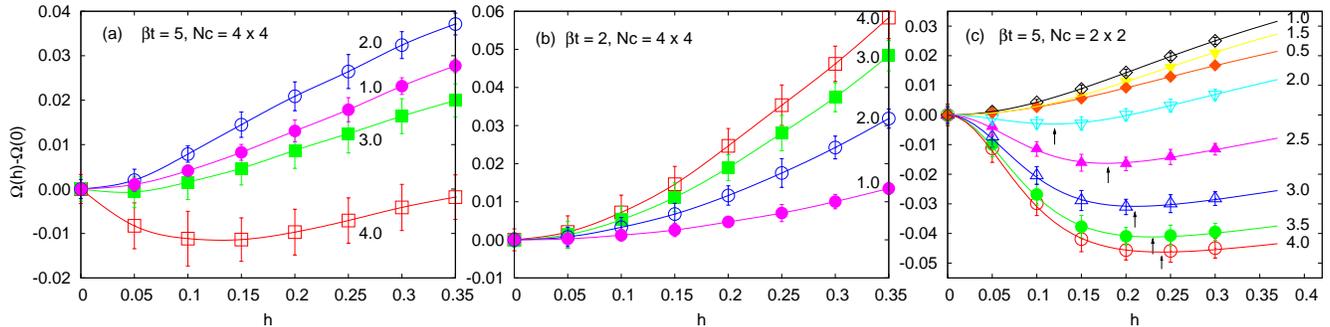}
\caption{Lattice grand potential $\Omega$ as a function of the variational parameter $h$ as obtained by embedding a $4\times 4$ cluster (a,b) and a $2\times 2$ cluster (c). Results for different interaction strengths $U/t$ as indicated in the figure. Note that the difference $\Omega(h) - \Omega(0)$ is plotted.
(a) $\beta t=5.0$, $N_c = 4 \times 4$, (b) $\beta t=2.0$, $N_c = 4 \times 4$, (c) $\beta t = 5.0$, $N_c = 2 \times 2$.
Symbols with statistical errors represent the numerical data. Lines are obtained by a spline interpolation and serve as a guide to the eye only in panel (a) and (b). Lines in panel (c) are from full ED calculations for comparison.
}
\label{Fig:Omega_h}
\end{figure*}

The general trend of $\Omega(T)$ is ruled by general thermodynamical relations. Its negative slope is given by the entropy, the second derivative corresponds to the specific heat divided by $T$. The linear high-temperature trend reflects the Hilbert-space dimension $4^N$ of the Hubbard model with $N$ sites since
\begin{equation}
\lim_{T\rightarrow\infty} S(T) = \ln 4^N = 2N \ln 2 
\end{equation}
for the entropy. At low temperatures, the results might be indicative of a quadratic behavior around $T=0$, corresponding to a linear entropy and a linear specific heat, as one would expect for a Fermi liquid. Note that there is hardly a change of the temperature trend when varying $U$. More definite statements, however, would require calculations at considerably lower $T$. 

Fig.\ \ref{Entropy} shows the entropy calculated from
\begin{equation}\label{Eq:Entropy}
S = \frac{E - \Omega - \mu N n}{T} \: , 
\end{equation}
as a function of temperature for two different interaction strengths. Here $n$ is the particle number per site. 
It is obvious that at low $T$ and for $U/t=4$ the results for $S$ are less accurate when compared with those for the grand potential or internal energy although the statistical error is reasonably small. Via Eq.\ (\ref{Eq:Entropy}), a small $T$ in the denominator enhances the systematic error of $E$ and $\Omega$ resulting from the cutoff $k_c$. To get a more satisfactory result, $k_{c}$ would have be to increased to include more diagrams in the CT-QMC. 

The entropy can be also obtained as $S = - \partial \Omega / \partial T$ from the grand potential.
There is no reason why the two ways for computing $S$ should give the same result within an approximate theory. Opposed to quantities like the particle number or the staggered magnetization (see Ref.\ \onlinecite{Aichhorn-2006}, for example), there is no internal consistency of a cluster-embedding approach with respect to the entropy. We have therefore explicitly checked for differences between both. 
Fig.\ \ref{Fig:Compare_S} shows the temperature dependence of the entropy at $U/t=1$ as obtained from Eq.\ (\ref{Eq:Entropy}) and from Fig.\ \ref{GrandPotential} by differentiation, respectively. 
Finite differences show up at intermediate temperatures. To a major extent, these are due to the error in calculating the temperature derivative from the discrete and small number of $\Omega$ values. 
This can be seen e.g.\ at temperature $T/t=2$ in the figure. There is, however, a small but significant remaining difference between both ways for computing $S$. This represents an intrinsic problem of the theory, and actually of any cluster-embedding approach. The observed inconsistency may serve as a measure for the error due to the finite cluster size since thermodynamically consistent results for $S$ can be expected strictly in the infinite-cluster limit only. 

Finally, Fig.\ \ref{Fig:Omega_h} demonstrates that our technique can in fact be used for the determination of variational parameters within the framework of the self-energy-functional theory.
The figure shows the lattice grand potential $\Omega$ as a function of the strength of the Weiss field $h$ (see Eq.\ \ref{eq:weiss}) for different $U$ and $T$. 
Let us concentrate on the results obtained for the $4\times 4$ cluster first.
For $U/t=1$, $U/t=2$ and $\beta t=5.0$ (panel a) we find a single stationary point at $h=0$. 
This is indicative of the paramagnetic phase at high temperatures and weak interaction.
For $U/t=4$ (panel a), the SFT grand potential clearly displays a minimum around $h=0.15$. 
This corresponds to antiferromagnetic order. 
Here we also get a non-zero value for the order parameter from Eq.\ (\ref{eq:m}).
Note that the variation of $\Omega$ with $h$ is small and comparable to the statistical error. 
This shows that $\beta t=5$ is close to the Neel temperature for $U/t=4$ (and for the given cluster size, see below).
As the trend of $\Omega(h)$ is symmetric with respect to $h$, the point $h=0$ represents another stationary point corresponding to the paramagnetic phase. 
The latter is metastable as $\Omega(h=0)$ is higher than $\Omega(h=0.15)$.

Fig.\ \ref{Fig:Omega_h}, panel (b) displays results for a higher temperature ($\beta t =2$).
Here we are again left with the paramagnetic phase for all $U/t$ only.
Obviously, the variation of the grand potential with $h$ is most pronounced for $U/t=4$ while is becomes more and more flat with decreasing interaction strength.
This is due to the fact that the $h$ dependence enters the self-energy functional via the self-energy only, i.e.\ $\Omega(h) = \Omega[\ff \Sigma(h)]$, and $\Omega[\ff \Sigma] \equiv 0$ in the non-interacting limit. 
For $\beta t = 5$ (panel a), this is different. 
Here, the trend of $\Omega(h)$ first becomes stronger (comparing $U/t=1$ with $U/t=2$) as explained above.
For stronger interactions, however, this mechanism has to compete with the tendency to form a minimum at a finite $h$.
This explains the non-monotonic trend in $U$ visible in panel a.

Due to the Mermin-Wagner theorem \cite{Mermin-1966} there is no long-range antiferromagnetic order in the two-dimensional Hubbard model at finite temperatures. 
Therefore, in a strict interpretation, the symmetry-broken phase obtained from the cluster-embedding approach has to be seen as a mean-field artifact.. 
In the context of the dynamical cluster approximation, for example, this has been studied extensively in Ref.\ \onlinecite{MJS+05}.
For our case, a non-zero value of optimal Weiss field actually indicates that the antiferromagnetic correlation length $\xi$ exceeds the linear size of the cluster, i.e.\ four sites.

This interpretation is corroborated by the comparison of the results from the $4\times 4$ cluster embedding with the corresponding ones obtained from a $2\times 2$ cluster (see panel c). 
Here we have also added the results using full ED as a cluster solver for comparison.
Antiferromagnetic (short-range) order at finite temperatures is build up with increasing interaction.
If, within a cluster mean-field approach, a symmetry-broken state is obtained once $\xi \sim \sqrt{N_c}$, one should therefore expect a lower critical interaction strength when reducing the cluster size $N_c$.
This is exactly what is found when comparing the results for $N_c=4\times 4$ (panel a) with those for $N_c=2\times 2$ (panel c). 
At $U/t=4$ the optimal value for $h$ for for the $2\times 2$ cluster ($h_{\rm opt}=0.24$) is larger than that for the $4\times 4$ cluster ($h_{\rm opt}=0.15$).
Furthermore, the difference $\Omega(h)-\Omega(0)$, which measures the stability of the antiferromagnetic order, is higher for the $2\times 2$ calculation as compared to the $4 \times 4$ one at the same $U/t$.

\section{Conclusions}

Besides the calculation of one-particle excitations, cluster-embedding approaches are able to provide detailled information on the thermodynamical properties of correlated lattice-fermion models, such as the Hubbard model. When combined with the quantum Monte-Carlo technique as a cluster solver, however, cluster DMFT schemes or the VCA cannot directly access a thermodynamical potential because at finite temperatures the entropy term in the free energy, for example, cannot be obtained by measuring an observable. On the other hand, a thermodynamical potential is of crucial importance for the construction of phase diagrams as it decides on the relative stability in situations where there are different competing phases, e.g.\ in the high-$T_c$ problem.

For the novel continuous-time QMC schemes, the present study has demonstrated that there is an elegant way to overcome this difficulty, namely by combining the CT-QMC approach with a quantum Wang-Landau technique. Employing a proper reweighting of the transition probabilities, it is easily possible to construct a flat distribution of the perturbation order $k$ up to an in principle arbitrarily high cutoff order. This allows for a direct calculation of the finite-$T$ partition function of the cluster. This has been demonstrated here for the weak-coupling variant of CT-QMC but can be generalized to other schemes. Note that modest changes of the QMC code are requried only. Another advantage of the Wang-Landau technique which is worth mentioning consists in the fact that the $U$ dependency of observables is obtained in a single simulation: Since the weight factors are available for any perturbation order separately, the $U$-dependence of observables is trivially given within the weak-coupling CT-QMC. Once the perturbation-order cutoff $k_c$ is suitably fixed for a calculation at interaction strength $U_0$, the same weight factors determine the entire $U$ dependence of the observables for $U\le U_0$. Note, however, that one cannot make use of this advantage within (cluster) DMFT, since the effective cluster problem is itself $U$ dependent. 

The lattice contribution to the total grand potential is obtained from the cluster Green's function and the lattice Dyson equation by a proper summation over Matsubara frequencies in the $\Tr \ln \ff G$-like terms. To obtain convergent results, the high-frequency asymptotics must be controlled carefully. 

Compared to full diagonalization of the cluster problem or to Lanczos-type approaches, the CT-QMC-Wang-Landau method can used for larger clusters with moderate computational effort. For a proof of principle, we have applied the VCA to the Hubbard model on a square lattice at half-filling and considered a single variational parameter only. A cluster with $4\times 4$ sites has been used for the embedding. This is clearly beyond the range that has been accessible in former VCA studies (at $T=0$) which were based on the Lanczos method and thereby restricted to clusters with about 12 sites at most. 

It is important to note, however, that one of the advantages of the QMC technique is that uncorrelated bath sites can be integrated out and can thus be included in the cluster reference system without any extra numerical cost. The calculations presented here have been done without bath sites. While this is sufficient to discuss the methodical aspects of the CT-QMC-Wang-Landau approach, future studies should be carried out by including a continuous bath, i.e.\ the VCA should be replaced by cellular DMFT (or a different cluster-DMFT variant) for finite-$T$ calculations based on QMC. This ensures an optimal description of the local quantum fluctuations on the cluster without extra cost. The implementation of the quantum Wang-Landau technique as well as the computation of the lattice contribution is identical to the VCA-based study shown here.

\acknowledgments
The authors would like to thank F.F. Assaad and E. Arrigoni for many useful discussions. Our numerical calculations were caried out at the LRZ-Munich Supercomputer center. The work was supported by the Bavarian Supercomputing Network KNOWIHR II and the Deutsche Forschungsgemeinschaft within the Forschergruppe FOR~538 and the Sonderforschungsbereich 668.

\bibliographystyle{apsrev}	
\bibliography{ref}
\end{document}